\documentclass[12pt]{article}
\usepackage{graphicx,amsmath}
\usepackage{units}

\newlength{\dinwidth}
\newlength{\dinmargin}
\def\lapproxeq{\lower .7ex\hbox{$\;\stackrel{\textstyle                                                    
<}{\sim}\;$}}                                                    
\def\gapproxeq{\lower .7ex\hbox{$\;\stackrel{\textstyle                                                    
>}{\sim}\;$}}                                                    
\def\be{\begin{equation}}                                                    
\def\ee{\end{equation}}                                                    
\def\bea{\begin{eqnarray}}                                                    
\def\eea{\end{eqnarray}}

\def\sh{\hat s}
\def\sh2{{\hat s}^2}

\begin{document}
                                                    
\titlepage                                                    
\begin{flushright}                                                    
IPPP/19/60  \\                                                    
\today \\                                                    
\end{flushright} 
\vspace*{0.5cm}
\begin{center}                                                    
{\Large \bf Colliding Pomerons}\\
\vspace*{1cm}
                                                   
V.A. Khoze$^{a,b}$, A.D. Martin$^a$ and M.G. Ryskin$^{a,b}$ \\                                                    
                                                   
\vspace*{0.5cm}                                                    
$^a$ Institute for Particle Physics Phenomenology, University of Durham, Durham, DH1 3LE \\                                                   
$^b$ Petersburg Nuclear Physics Institute, NRC Kurchatov Institute, Gatchina, St.~Petersburg, 188300, Russia

\vspace*{1cm}

\begin{abstract}

  We recall the main properties of inclusive particle distributions expected for Pomeron-proton and Pomeron-Pomeron interactions. Due to the small size of the Pomeron we expect  larger transverse momenta of secondaries and a smaller probability of Multiple Interactions, that is a narrower multiplicity distribution.
We propose to compare the spectra of secondaries produced in the Pomeron
and the proton interactions in terms of the Feynman $x_F$ variable. The main difference should be observed for a relatively large $x_F$, that is near the edge of rapidity gaps. Such data offer the opportunity to illuminate the properties of the `soft' or `Regge' Pomeron, which drives the minimum-bias type of events in high energy $pp$ interactions.  Besides this, there should be a good opportunity to observe a glueball in the Pomeron fragmentation region.
\end{abstract}
\end{center}  

\vspace*{1cm}

\section{Experimental possibilities}
In a recent paper~\cite{Luk} the inclusive distribution of identified particles produced in Single Diffractive Dissociation (SD) $pp\to p+X$ and in Central Diffractive  (CD) $pp\to p+X+p$ processes were studied with
the STAR detector at RHIC in
  proton-proton collisions at $\sqrt s=200$ GeV. Here $X$ denotes the diffractively produced system. The SD and CD events were selected by observing in the STAR Roman Pot system(s)
 leading proton (or protons) which carry a large fraction, $x_L$, of the beam momentum. We denote  
$x_L=1-\xi$. 

\begin{figure} [t]
\begin{center}
\includegraphics[trim=0.0cm 0cm 0cm 0cm,scale=0.7]{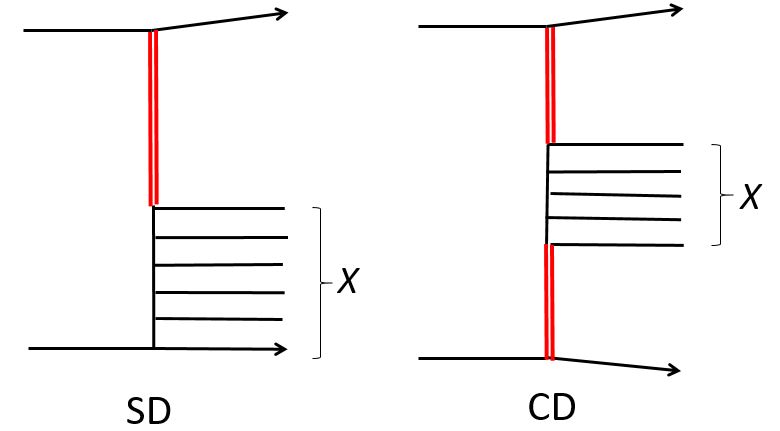}
\caption{\sf Sketches of Single Diffractive (SD) $pp\to p+X$ and Central Diffractive (CD) $pp\to p+X+p$ processes, which respectively arise in Pomeron-proton and Pomeron-Pomeron collisions.
}
\label{fig:1}
\end{center}
\end{figure}

Note that after the leading proton(s) with a large $x_L$ close to 1 are detected we have rather small remaining energy to produce the new secondaries. Therefore,
these new secondaries (system $X$) are separated from the leading proton(s) by Large Rapidity Gap(s) (LRG) with size\footnote{For $pp\to p+X+p$ the mass $M$ of the centrally produced system is given by $M^2=s(1-x_{L1})(1-x_{L2})$ with gap sizes $\Delta y_i\simeq -\ln(1-x_{Li})$.} $\Delta y\simeq \ln(1/\xi)$. Since the interaction across the LRG is provided by the Pomeron exchange
such events can be interpreted as the result of Pomeron interaction with a proton (SD) or a Pomeron (CD) target. The processes are illustrated in Fig.~\ref{fig:1}.

Unfortunately, the $\xi$ interval selected in the STAR measurement \cite{Luk} was rather large, 
from 0.02 to 0.4. In such a case we cannot be sure that we deal with pure {\em diffractive} events. There could be an admixture of secondary Reggeons. Moreover, moderate size rapidity gaps can be formed in  the usual inelastic process by  fluctuations at the hadronization stage (see~\cite{Frank}). It would be more informative to perform an analogous study
at  RHIC or at the LHC with a smaller $\xi<0.05$. Of course in the case of RHIC by selecting such a small $\xi$ we deal with a relatively low proton-Pomeron energy; namely $M=\sqrt{\xi S_{pp}}=28-45$ GeV  (and $M=\xi\sqrt{S_{pp}}=4-10$ GeV for CD case) for the RHIC energy $\sqrt{S_{pp}}=200$ GeV. However 
it will be interesting to compare the properties of the Pomeron and the proton interactions at so small energies as well. It is not necessary that the Pomeron interaction must be described by the Pomeron exchange only. The contribution of secondary Reggeons and/or the Pomeron-Pomeron fusion to $s$-channel resonances is not suppressed a priori. 

The aim of the present note is to recall what we expect for the Pomeron-proton and Pomeron-Pomeron inelastic interactions (see also~\cite{9,11}).

\section{Theoretical expectations}
 First,
 the Pomeron has a small transverse size. In comparison with the proton radius $\sim 1$ fm the Pomeron size is $\sim 0.1$ fm. This is indicated by the small value of the Pomeron trajectory slope $\alpha'_P\leq 0.25$ GeV$^{-2}$ (see e.g.~\cite{DL,KMR-slope,GLM})~\footnote{It was shown long ago in terms of the multiperipheral models~\cite{FC} and in terms of the parton cascade~\cite{Gr} that the value of $\alpha'\propto 1/k^2_t$ where $k_t$ is the typical transverse momentum of the partons (or of $t$-channel propagators in the case of multiperipheral models). Simultaneously this value of $k_t$ determines the size of the bound system which forms the Regge pole (Pomeron).}   and the very small (consistent with zero) $t$-slope of the triple-Pomeron vertex (see e.g.~\cite{KKPT,FF,Luna})~\footnote{In these papers the triple Pomeron vertex was extracted fitting to rather old CERN-ISR data (Tevatron data was included in~\cite{Luna}). However this energy was sufficient to separate the triple-Pomeron contribution, and since the vertex occupies a small rapidity interval we can use the obtained results at larger  energies; in particular for the LHC region.}  . Therefore, we expect a larger mean transverse momenta, $p_t$, and a broader $p_t$ distribution of the secondaries produced near the edge of LRG. Some indication in favour of this can be seen in Fig.2 of~\cite{Luk} where  in comparison with the PYTHIA 8 Monte Carlo simulations the particle density increases with $p_t$.  The `data/MC' ratio exceeds 1 and reaches about 2 for $p_t>1$ GeV.

Next, the `effective' Pomeron-proton cross section is rather small. For example, from the triple-Regge analysis~\cite{Luna} the ratio of the triple-Pomeron coupling, $g_{3P}$, to the proton-Pomeron coupling, $g_N$ is about $g_{3P}/g_N=\lambda=0.2$. 
Thus, due to the small Pomeron size and a smaller `effective' Pomeron-proton cross section the probability of an additional soft interaction is strongly suppressed in the Pomeron-proton and the Pomeron-Pomeron cases in comparison with proton-proton interactions. In other words, we expect a much smaller probability of Multiple Interactions (MI) for SD and CD events. This means we obtain  narrower multiplicity distributions when selecting SD and/or CD processes.

Recall also that the Pomeron consists mainly of gluons  and so the Pomeron is essentially a singlet with respect to the flavour SU(3) group. Therefore, it would be interesting to observe in the Pomeron fragmentation region (close to the edge of the LRG) the presence of $\eta$ and $\eta'$ mesons. Since $\eta'$ is almost a singlet of flavour SU(3)
and contains a large gluon component we may expect that the Pomeron
fragmentation region will be enriched by $\eta'$ mesons. Besides this, there should be a good chance to observe $0^{++}$ and $2^{++}$ glueballs in the Pomeron fragmentation region.

\section{Useful observables}
As was mentioned in~\cite{9,11} it would be interesting to compare the spectra of secondaries produced in the Pomeron-proton and proton-proton interactions~\footnote{It would be best to compare the spectra at the same energy $s(pp)=s(p$-${\rm Pomeron})=\xi S_{pp}$ or, for the CD case, $s(pp)=s({\rm Pomeron}$-${\rm Pomeron})=\xi_1\xi_2 S_{pp}$, where $S_{pp}$ is the  energy squared of the initial interaction from which the events with leading (large $x_L$) protons are selected.}. In more detail, we note that it would be appropriate to study the distributions in terms of the Feynman $x_F$ variable. $x_F$ is the Pomeron (incoming beam) momentum fraction carried by the secondary particle. That is, we would prefer to work with quantities such as
\begin{equation}
\label{xF}
\frac{dN}{dx_F}=\frac 1{\sigma_{\rm inel}}\frac{d\sigma}{dx_F}~~~\;\;\;\mbox{and}~~~\;\;\;
\frac{dN}{dx_Fdp^2_t}=\frac 1{\sigma_{\rm inel}}\frac{d\sigma}{dx_Fdp^2_t}\ .
\end{equation}
According to Feynman $x_F$ scaling~\cite{F} (or Limiting Fragmentation~\cite{LF}) these distributions (\ref{xF}) should depend 
weakly on energy, but at moderate $x_F$ the spectra should depend on the quantum numbers of the incoming beam particle. In particular, it would be interesting to compare the ratios, such as $K/\pi$, of kaons and pions (and/or other ratios) depending on $p_t$ and $x_F$. In spite of the fact that in the soft regime the particle ratios are driven mainly by hadronization, in Pomeron interactions we may expect a  smaller suppression of strange quarks and hence a bit (10-20\%) larger ratio $K/\pi$.

Next, we expect that the mean $p_t$ of (each sort of) secondaries measured at not too small values of $x_F$ in Pomeron-proton collisions should be larger than that in proton-proton interactions at the same value of $x_F$.

\section{Possible measurements at the LHC}
These measurements can be performed at the LHC using the forward proton detectors \cite{FPD,CMS,FWG} or just by selecting events with 
 large rapidity gaps.
\begin{equation}
\label{2}
pp\to p^*~+~X\ ,
\end{equation}
\begin{equation}
\label{3}
pp\to p^*~+~X~+~p^*\ ,
\end{equation}
where `$+$' indicates the presence of a LRG and $p^*$ allows for accompanying particles to the forward going protons.
The expected cross sections are rather large
\begin{equation}
\frac{d\sigma}{dy}\simeq 0.5~\mbox{mb}
\end{equation}
for the SD processes (\ref{2}) with rapidity $y$ ~\cite{SD} relevant to the edge of the LRG, and
\begin{equation}
\frac{\xi_1\xi_2d\sigma}{d\xi_1d\xi_2}\sim 1-4~\mu\mbox{b}
\end{equation}
for the case of CD~\cite{CD}. Therefore the corresponding studies can be performed during short low luminosity runs. In order to collect, say, 100 thousands of SD events in the interval of $\xi=0.02-0.06$ it will be sufficient
to have an integrated luminosity of about $0.2$ nb$^{-1}$. For the case of CD the luminosity of 20 nb{$^{-1}$ will provide about 10000 events (in the same $\xi$ interval).

Note that using the forward proton detectors at the LHC in order to study Pomeron fragmentation we have to select very small $\xi\sim 10^{-3}$. Otherwise  in the central detector with $|\eta|<2.5$ we will already observe the 
central plateau  region of the Pomeron-proton secondary distribution where the specific properties of the Pomeron (small size, vacuum quantum numbers, gluon dominance) will not manifest themselves so prominently.
 Nevertheless in comparison with proton-proton collisions (at the same  proton beam energy) we may expect about a twice smaller (due to the suppression of Multiple Interactions for LRG events) particle density, $dN/dy$, observed in the central detector and correspondingly - a twice smaller multiplicity and dispersion (width) of the $\sigma_N$ distribution. It is more complicated to make the prediction for the  $p_t$ spectra of secondaries coming from this {\em soft} interaction. Here the expected difference is smaller since the $p_t$ distributions are driven mainly by the dynamics of hadronization. Very roughly we expect almost the same (may be a little bit smaller) mean $\langle p_t\rangle $ as that in the normal inelastic interaction 
 at the same proton beam energy but about 10 - 20\% larger than that at measured
 at $\sqrt{s_{pp}}=M_{\rm Pomeron -proton}$ for SD case or  $\sqrt{s_{pp}}=M_{\rm Pomeron -Pomeron}$ for CD events.
 
 On the other hand, by selecting Pomeron interactions via the `gap-veto' in a low luminosity run (that is selecting events with LRGs) we  have the possibility to measure features of Pomeron interactions  just in the Pomeron fragmentation region near the edge of rapidity gaps. The events where the edge of the LRG can be seen in the central detector  look as follows: in some rapidity interval, say $\eta>\eta_0$ we observe a rather large density of secondaries while at $\eta<\eta_0$ the detector is empty~\footnote{Here we neglect the possibility of pile-up events. In the presence of `pile-up' the requirement ``tracks at $\eta>\eta_0$  and no tracks with $\eta<\eta_0$"  should be applied for one particular vertex (that is for a single event).}. The observed, in an individual event, value of $\eta_0$ determines the position of LRG edge.   Such data near the gap edge, that is coming from the Pomeron fragmantation region, will tell us more about the properties of the `soft' or `Regge' Pomeron, which drives the minimum-bias events in $pp$ interactions.

It would be also interesting to study the Bose-Einstein correlations
in order to evaluate the size of the region from which the secondaries are produced in Pomeron-proton and/or Pomeron-Pomeron interactions~\cite{bec}. In comparison with the proton-proton case we expect a smaller size due to the small transverse size of the Pomeron and  a smaller probability of MI (i.e. of multi-Pomeron exchange).

\section*{Acknowledgments}
 We thank Valery Schegelsky and Marek Tasevsky for the discussion.
 MGR thanks the IPPP at the University of Durham for hospitality.
VAK acknowledges support from a Royal Society of Edinburgh Auber award.

G.~Aad {\it et al.} [ATLAS Collaboration],
  Eur.\ Phys.\ J.\ C {\bf 72}, 1926 (2012)
  [arXiv:1201.2808 [hep-ex]];\\
  ATLAS collaboration,
ATLAS-CONF-2019-012;\\
V.~Khachatryan {\it et al.} [CMS Collaboration],
  Phys.\ Rev.\ D {\bf 92},
   012003 (2015)
[arXiv:1503.08689 [hep-ex]].

\bibitem{CD} 
 M.~G.~Ryskin, A.~D.~Martin and V.~A.~Khoze,
  Eur.\ Phys.\ J.\ C {\bf 60} (2009) 249
[arXiv:0812.2407 [hep-ph]] (see Fig. 12).

\bibitem{bec}
 V.~A.~Khoze, A.~D.~Martin, M.~G.~Ryskin and V.~A.~Schegelsky,
  Eur.\ Phys.\ J.\ C {\bf 76} (2016)  193
  [arXiv:1601.08081 [hep-ph]].


\thebibliography{}
\bibitem{Luk} 
L.~Fulek (for the STAR Collaboration)
  arXiv:1906.04963 [hep-ex].
\bibitem{Frank}
V.~A.~Khoze, F.~Krauss, A.~D.~Martin, M.~G.~Ryskin and K.~C.~Zapp,
  Eur.\ Phys.\ J.\ C {\bf 69}, 85 (2010)
[arXiv:1005.4839 [hep-ph]].


\bibitem{9}
M.~G.~Ryskin, A.~D.~Martin, V.~A.~Khoze and A.~G.~Shuvaev,
J.\ Phys.\ G {\bf 36} (2009) 093001
  [arXiv:0907.1374 [hep-ph]] (see Section 7).

\bibitem{11}
M.~G.~Ryskin, A.~D.~Martin and V.~A.~Khoze,
J.\ Phys.\ G {\bf 38}, 085006 (2011)
 [arXiv:1105.4987 [hep-ph]] (see Sections 5.3, 5.4, 7).

\bibitem{DL} 
A.~Donnachie and P.~V.~Landshoff,
  Nucl.\ Phys.\ B {\bf 231}, 189 (1984).

\bibitem{KMR-slope} 
V.~A.~Khoze, A.~D.~Martin and M.~G.~Ryskin,
  Eur.\ Phys.\ J.\ C {\bf 73}, 2503 (2013)
  [arXiv:1306.2149 [hep-ph]].

\bibitem{GLM} 
E.~Gotsman, E.~Levin and U.~Maor,
  Int.\ J.\ Mod.\ Phys.\ A {\bf 30} (2015) no.08,  1542005
  [arXiv:1403.4531 [hep-ph]].
\bibitem{FC} E.L. Feinberg and D.S. Chernavski, Usp. Fiz. Nauk {\bf 82} (1964) 41.  
\bibitem{Gr} V.N. Gribov, Yad. Fiz. {\bf 9} (1969) 640, Sov.J.Nucl.Phys. {\bf 9} (1969) 369
\bibitem{KKPT} 
  A.~B.~Kaidalov, V.~A.~Khoze, Y.~F.~Pirogov and N.~L.~Ter-Isaakyan,
 Phys.\ Lett.\  {\bf 45B} (1973) 493.

\bibitem{FF}
 R.~D.~Field and G.~C.~Fox,
 Nucl.\ Phys.\ B {\bf 80}, 367 (1974).


\bibitem{Luna}
 E.~G.~S.~Luna, V.~A.~Khoze, A.~D.~Martin and M.~G.~Ryskin,
Eur.\ Phys.\ J.\ C {\bf 59} (2009) 1
[arXiv:0807.4115 [hep-ph]].

\bibitem{F} 
 R.~P.~Feynman,
  Phys.\ Rev.\ Lett.\  {\bf 23} (1969) 1415.


\bibitem{LF}
J.~Benecke, T.~T.~Chou, C.~N.~Yang and E.~Yen,
  Phys.\ Rev.\  {\bf 188}, 2159 (1969).


\bibitem{FPD} 
L.~Adamczyk {\it et al.},
  CERN-LHCC-2015-009, ATLAS-TDR-024.


\bibitem{CMS} 
M.~Albrow {\it et al.} [CMS and TOTEM Collaborations],
CERN-LHCC-2014-021, TOTEM-TDR-003, CMS-TDR-13.

\bibitem{FWG} 
K.~Akiba {\it et al.} [LHC Forward Physics Working Group],
  J.\ Phys.\ G {\bf 43}, 110201 (2016)
  [arXiv:1611.05079 [hep-ph]].

\bibitem{SD} 
\end{document}